\documentclass[aps,twocolumn,preprintnumbers,amsmath,amssymb,superscriptaddress]{revtex4}
\usepackage[per-mode=symbol]{siunitx}
\usepackage{graphicx}
	\graphicspath{{figures/}}
\usepackage{csquotes}
\usepackage[hidelinks]{hyperref}
\usepackage{cleveref}
\Crefname{figure}{Fig.\hspace{-.2em}}{Figs.\hspace{-.2em}}
\begin{document}

\title{Photo\-electron spectroscopy of laser-dressed atomic helium}

\author{Severin Meister}
\email{severin.meister@mpi-hd.mpg.de}
\affiliation{Max-Planck-Institute for Nuclear Physics, 69117 Heidelberg, Germany}
\author{Aaron Bondy}
\affiliation{Department of Physics and Astronomy, Drake University, Des Moines, IA 50311, USA}
\affiliation{Department of Physics, University of Windsor, Windsor, Ontario N9B 3P4, Canada}
\author{Kirsten Schnorr}
\author{Sven Augustin}
\affiliation{Paul Scherrer Institute, 5232 Villigen, Germany}
\author{Hannes Lindenblatt}
\author{Florian Trost}
\affiliation{Max-Planck-Institute for Nuclear Physics, 69117 Heidelberg, Germany}
\author{Xinhua Xie}
\affiliation{Paul Scherrer Institute, 5232 Villigen, Germany}
\author{Markus Braune}
\author{Rolf Treusch}
\affiliation{Deutsches Elektronen-Synchrotron, 22607 Hamburg, Germany}
\author{Nicolas Douguet}
\affiliation{Department of Physics, Kennesaw State University, Marietta, GA 30060, USA}
\author{Thomas Pfeifer}
\affiliation{Max-Planck-Institute for Nuclear Physics, 69117 Heidelberg, Germany}
\author{Klaus Bartschat}
\affiliation{Department of Physics and Astronomy, Drake University, Des Moines, IA 50311, USA}
\author{Robert Moshammer}
\affiliation{Max-Planck-Institute for Nuclear Physics, 69117 Heidelberg, Germany}

\begin{abstract} 
Photo\-electron emission from excited states of laser-dressed atomic helium 
is analyzed with respect to laser intensity-dependent excitation energy shifts and angular distributions. 
In the two-color XUV (exteme ultra\-violet) -- IR (infrared) measurement, the XUV photon energy is 
scanned between \SI{20.4}{\electronvolt} and the ionization threshold at \SI{24.6}{\electronvolt}, 
revealing electric dipole-forbidden transitions for a temporally overlapping IR pulse 
($\sim\!\SI{e12}{\watt\per \centi\meter\squared}$). The interpretation of the experimental results is 
supported by numerically solving the time-dependent Schr\"odinger equation in a 
single-active-electron approximation.

\end{abstract}

\maketitle


Photo\-electron spectroscopy is a powerful technique to obtain compositional and structural 
information about matter and to investigate light$-$matter interactions in general. It has 
been successfully employed and continuously developed over many decades in atomic and molecular 
physics~\cite{1stPAD}. Photo\-electrons carry information about the electronic bound and continuum 
states of the corresponding atom, as well as information about the absorbed and emitted photons. 

With the advent of intense optical lasers, multi\-photon absorption in atoms and 
molecules became feasible, enabling the observation of a variety of new phenomena, e.g., 
multiphoton excitation microscopy~\cite{MPEmicroscpoy}, 
resonance-enhanced multiphoton ionization (REMPI)~\cite{REMPI}, Doppler-free two-photon 
spectroscopy~\cite{DopplerFree1st,DopplerFree}, and high-harmonic generation 
(HHG)~\cite{HHGCork,HHGLewenstein}, to name just a few. One step further in the 
investigation of light$-$matter interactions is the implementation 
of two-color ionization and excitation schemes, which reveal laser-induced continuum 
structures~\cite{LICS_Exp_81} and light-induced structures 
(LIS)~\cite{LISLeone2012}. In the former case, the dressing laser field couples 
bound states to the continuum, giving rise to a resonant 
structure~\cite{LICS_Review_90,LICS_multiphot_82,LICS_Propose_76,LICS_early_Theo}. 
In the latter case, the ground state is coupled to excited states beyond the 
one-photon allowed dipole transition via two-color 
photo-excitation~\cite{Chini2013,Reduzzi_ortho_LIS,LIS_Molecule,LIS_intensity}.

In this Letter, we report the use of XUV (extreme ultra\-violet) radiation with 
tunable wavelength provided by the free-electron laser in Hamburg (FLASH) in combination 
with a synchronized IR laser to obtain a detailed picture of excited states in 
laser-dressed atomic helium. The XUV photon energy is scanned over the $1snp\,^1\!P$ 
Rydberg excitation-series to a value just below the ionization threshold. The 
superimposed IR pulses (\SI{800}{\nano \meter} wavelength) arrive with a freely 
adjustable time delay with respect to the excitation pulses. Their intensity is 
too low to ionize He in its ground state, but strong enough to ionize 
it from excited states that are temporarily reached via a combination of XUV and 
IR photons. The ionization yield and angular distributions are analyzed as a function 
of the XUV photon energy, the IR time-delay, and the IR intensity. In the case of 
temporally overlapping pulses, by absorption of one XUV and one or more IR photons, 
one electron is lifted from the ground into a continuum state through laser-dressed 
excited states, including those that are not accessible by pure single-photon excitation. 
The interpretation of the experimental results is supported by numerical calculations 
based on the time-dependent Schr\"odinger equation (TDSE) within the single-active 
electron (SAE) approximation. 

This two-color scheme has the clear advantage over, e.g., single-color REMPI setups, 
that the dominant contribution to the excitation energy is delivered by only one XUV 
photon, and hence the laser intensity can be kept low. Therefore, field-induced changes 
to energy levels and fragmentation are minimized.  This opens up precision-spectroscopic 
studies of atoms and molecules under less-perturbing conditions. 



The experiment was carried out with the reaction microscope (REMI) endstation~\cite{REMIexp,REMItech} 
at the free-electron laser (FEL) FLASH2~\cite{FLASH2_start,Ackermann2007}. FLASH2 features 
variable-gap undulators that allow to quickly change the photon wavelength~\cite{FLASH2_tunability} 
over a broad spectral range. During the measurements, the XUV photon 
energy was scanned in steps of \SI{0.2}{\electronvolt} from \SI{20.4}{\electronvolt} 
to just below the ionization threshold of atomic helium at 
\SI{24.6}{\electronvolt}~\cite{Helevel}. The FEL pulse-length was about \SI{40}{\femto\second} 
full-width at half maximum (FWHM) in intensity, and the pulse energy (\(<\SI{10}{\nano\joule}\)) 
was reduced to a level such that two-XUV-photon absorption in He can be neglected. 
Synchronized, but with a timing jitter of several ten femtoseconds, the IR probe-laser 
(\SI{800}{\nano\meter}) was superimposed collinearly with the XUV beam. In order to ensure optimal 
temporal overlap with the FEL pulses, an IR pulse-duration of about \SI{90}{\femto\second} FWHM was 
chosen. The IR pulse energy and the focusing conditions were adjusted such that intensities in the 
order of up to \SI{e13}{\watt\per\centi\meter\squared} were reached in the target. With a diameter 
of about \SI{30}{\micro\meter} the focus of the IR beam was significantly wider than the FEL focus 
($\approx$ \SI{10}{\micro\meter}). XUV and IR radiation were linearly polarized and aligned parallel 
to each other. Both beams were focused onto a dilute supersonic gas jet of atomic helium in the 
center of the REMI, which is equipped with two time- and position-sensitive 
detectors~\cite{REMI2003} to collect all charged fragments (electrons and ions) within the full 
\(4\pi\) solid angle. Time-of-flight and position information is used to retrieve 
the particles' momentum vectors at the time of ionization.

During the XUV photon-energy scans, the FEL pulse energy and the FEL beam diameter change slightly. 
Together with the energy-dependent absorber-foil transmissions and mirror 
reflectivities, this leads to small variations in the photon flux. 
In our analysis, these effects are corrected by normalizing the data for each XUV energy 
with the simultaneously recorded yield of H\(_2^+\) ions, which stem from a constant and weak 
background of H\(_2\) gas in the REMI chamber. The ionization cross section of H\(_2\) 
was taken from~\cite{H2CrossSec}.

In the experiment two sets of data were taken, one with an IR intensity of approximately
$I_{\rm high} \approx$ \SI{8e12}{\watt\per\centi\meter\squared} and one with 
$I_{\rm low} \approx$ \SI{1e12}{\watt\per\centi\meter\squared}. 
We note that the temporal jitter between FEL and IR, which
is comparable to the IR pulse duration, leads effectively to a lowering of the average
IR intensity for the case of overlapping pulses. The influence of this imperfect overlap 
of both pulses increases with the IR intensity in the same way as the contribution of 
non\-linear multi-photon transitions increases. Therefore, in the comparison 
to theory, a smaller difference in intensity between the low and high IR intensity case 
was chosen in our calculations in order to mimic the corresponding experimental conditions.

\begin{figure}[t]
	\centering
	\includegraphics[width=0.45\textwidth]{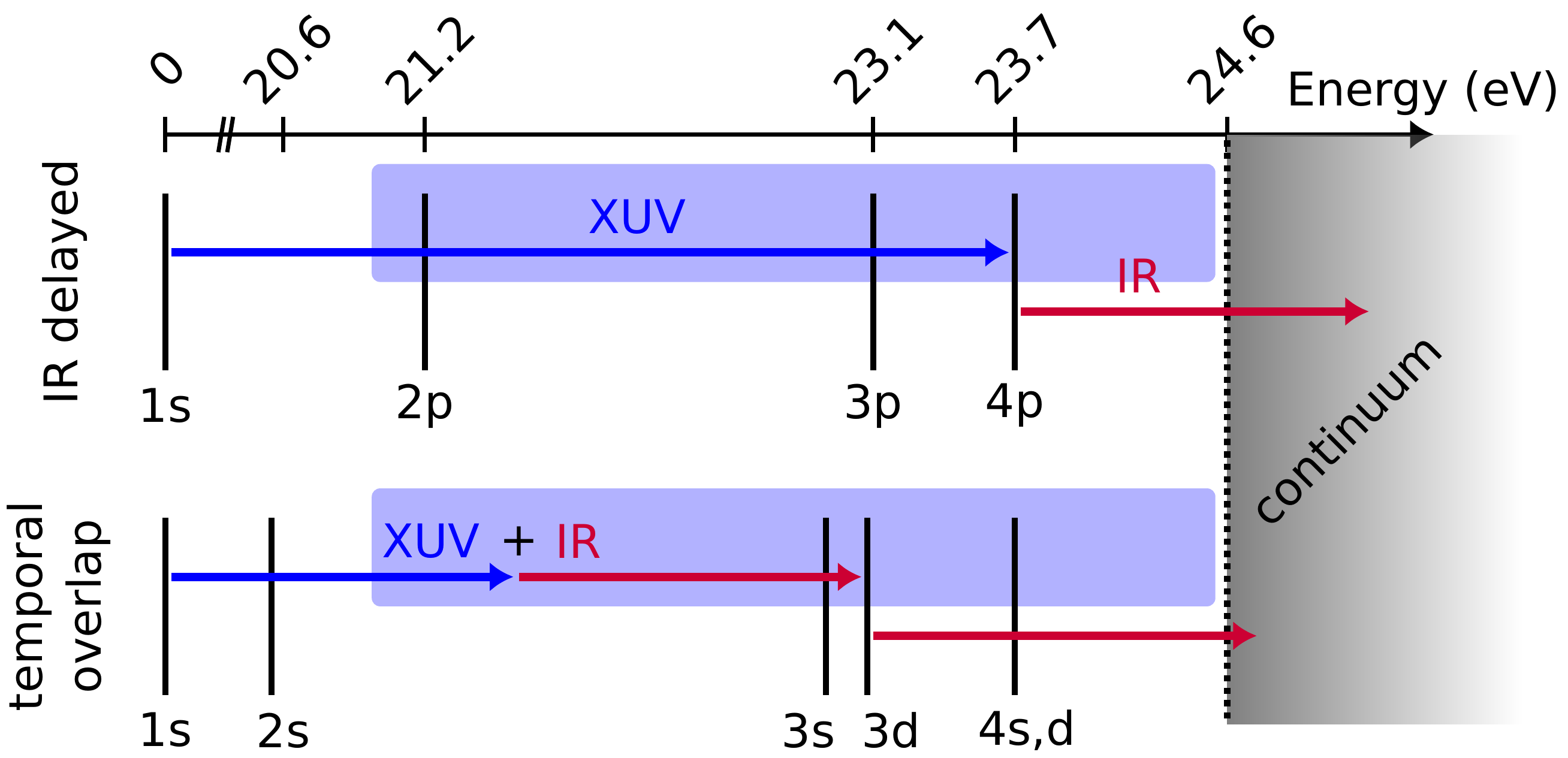}
	\caption{Electronic energy level scheme of helium with different ionization pathways
	         through intermediate excited states. 
		The XUV energy scanning range is indicated by the blue box.} \label{Levels}
\end{figure}

The theoretical part of this study is based on numerically solving the TDSE in the SAE approximation. 
Electrons are assumed to be non-interacting, while the ground state is effectively described as a 
$1s1s'\ ^1\!S$ state, where the $1s$ is close to the He$^+$ orbital and the $1s'$ is treated like a 
valence orbital. As always in theory, the binding energies of the $n\ell$ valence electrons are not 
exact. While $1s n \ell$ Rydberg states with angular momenta $\ell \ge 2$ have very accurate binding 
energies, this is not quite the case for $p$-electrons and particularly for $s$-electrons, due to 
the small or missing centrifugal barrier. Since excitation energies are measured from the ground state, 
much of the remaining discrepancies are due to the binding energy of the $1s'$ orbital.  

Specifically, we used the same one-electron potential as Birk {\it et al.}~\cite{Birk2020},
\begin{equation}
V(r) = -\frac{1}{r} - \left(\frac{1}{r} + 1.3313\right) \,\exp(-3.0634\,r),
\end{equation}
where $r$ is the distance from the nucleus, to calculate the valence orbitals. The difference of excitation 
energies compared to the recommended excitation from the NIST database~\cite{Helevel}, is less than 
0.2$\,$eV even in the worst-case scenario and does not alter the essential conclusions presented below. 
We will sometimes omit the inner $1s$ electron to simplify the notation, keeping in mind that 
only two-electron singlet spin states are accessible, since spin-forbidden transitions are negligible.

\begin{figure}[b]	
	\includegraphics[width=0.45\textwidth]{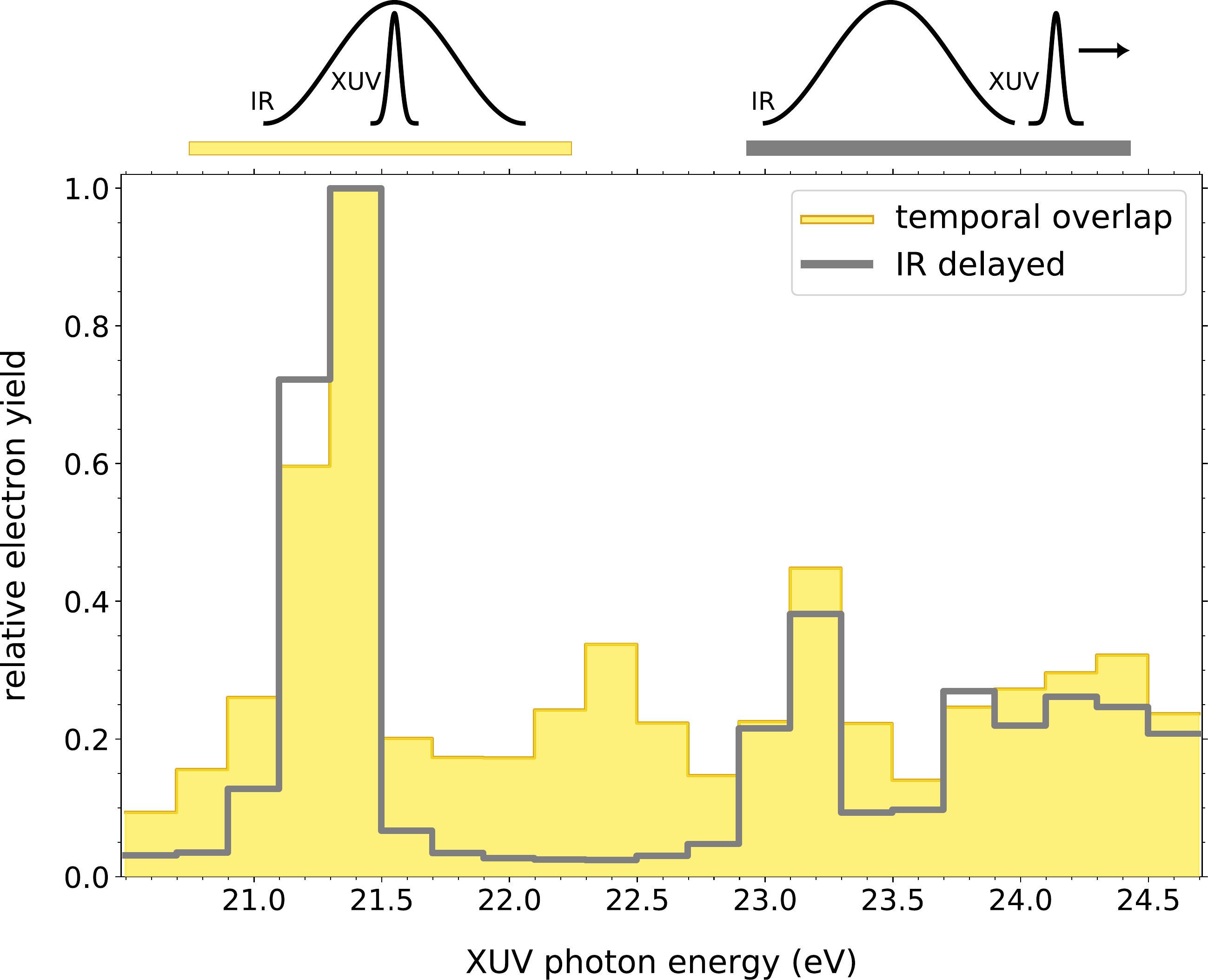}
	\caption{Photo\-electron yield measured for two different delays between the XUV and IR pulses. 
		Both curves are normalized to their maximum yield, and statistical 
		error bars are much smaller than the line thickness.\label{RelWeightDelay}}
	
\end{figure}

The laser parameters were chosen according to the available knowledge regarding the actual experimental 
conditions. The XUV pulse duration was taken as \SI{40}{\femto\second} (FWHM value of a peak intensity 
of \SI{1e10}{\watt/cm^2} with a Gaussian envelope) and the IR pulse duration as \SI{80}{\femto\second}. 
While the XUV photon energy was varied over a range in steps of \SI{0.05}{\electronvolt}, the central 
IR photon energy was held fixed at \SI{1.55}{\electronvolt} (\SI{800}{\nano\meter}). Since both beams 
are linearly polarized along the same direction, the initial state can be propagated very efficiently 
and accurately. Specifically, we used an updated version of the code described by Douguet 
{\it et al.}~\cite{PhysRevA.93.033402}.

Over the XUV scanning range, the helium atom can be excited from the $1s^2\,^1\!S$ ground state to 
a $1snp\,^1\!P$ excited state for specific XUV photon energies, according to the electric-dipole selection 
rules. The excited atom can be ionized by absorbing one or more IR photons 
($\hbar \omega = \SI{1.55}{\electronvolt}$) of a subsequent laser pulse, promoting the weakly bound 
electron into the continuum. This mechanism is depicted in an energy-level scheme in the upper part 
of \Cref{Levels}.

Experimental data for the corresponding process are shown in \Cref{RelWeightDelay},
where the yield of photo\-electrons is plotted 
against the XUV photon energy for a non\-overlapping temporally delayed IR pulse
with intensity $I_{\rm high}$. Clearly visible are the yield enhancements for XUV energies 
that match the $1snp\,^1\!P$ excitation energies in helium ($2p$ at \SI{21.2}{\electronvolt}, 
$3p$ at \SI{23.1}{\electronvolt}, $4p$ at \SI{23.7}{\electronvolt} etc.~\cite{Helevel}). 
Also shown in \Cref{RelWeightDelay} is the photo\-electron yield for XUV and IR pulses in 
temporal overlap (yellow distribution). Compared to delayed ionization, the 
\(^1\!P\)-excitation peaks remain while an additional maximum appears around \SI{22.4}{\electronvolt}. 
This feature was observed and described in transient-absorption measurements as a 
LIS~\cite{LISLeone2012,LISmolecule,Reduzzi_ortho_LIS,LIS_2011_Theo}. As XUV and IR 
radiation are simultaneously present, the helium atom can undergo dipole-forbidden (for single photons) 
transitions, provided the XUV photon absorption is accompanied by the absorption or emission of IR photons 
of the dressing laser field. In the simplest and dominant case, one XUV photon and one IR photon combined 
drive $1s^2 \rightarrow 1sns$ or $1s^2 \rightarrow 1snd$ transitions. By absorbing additional IR photons, 
the excited atom is ionized.

\begin{figure}[b]
	\includegraphics[width=0.45\textwidth]{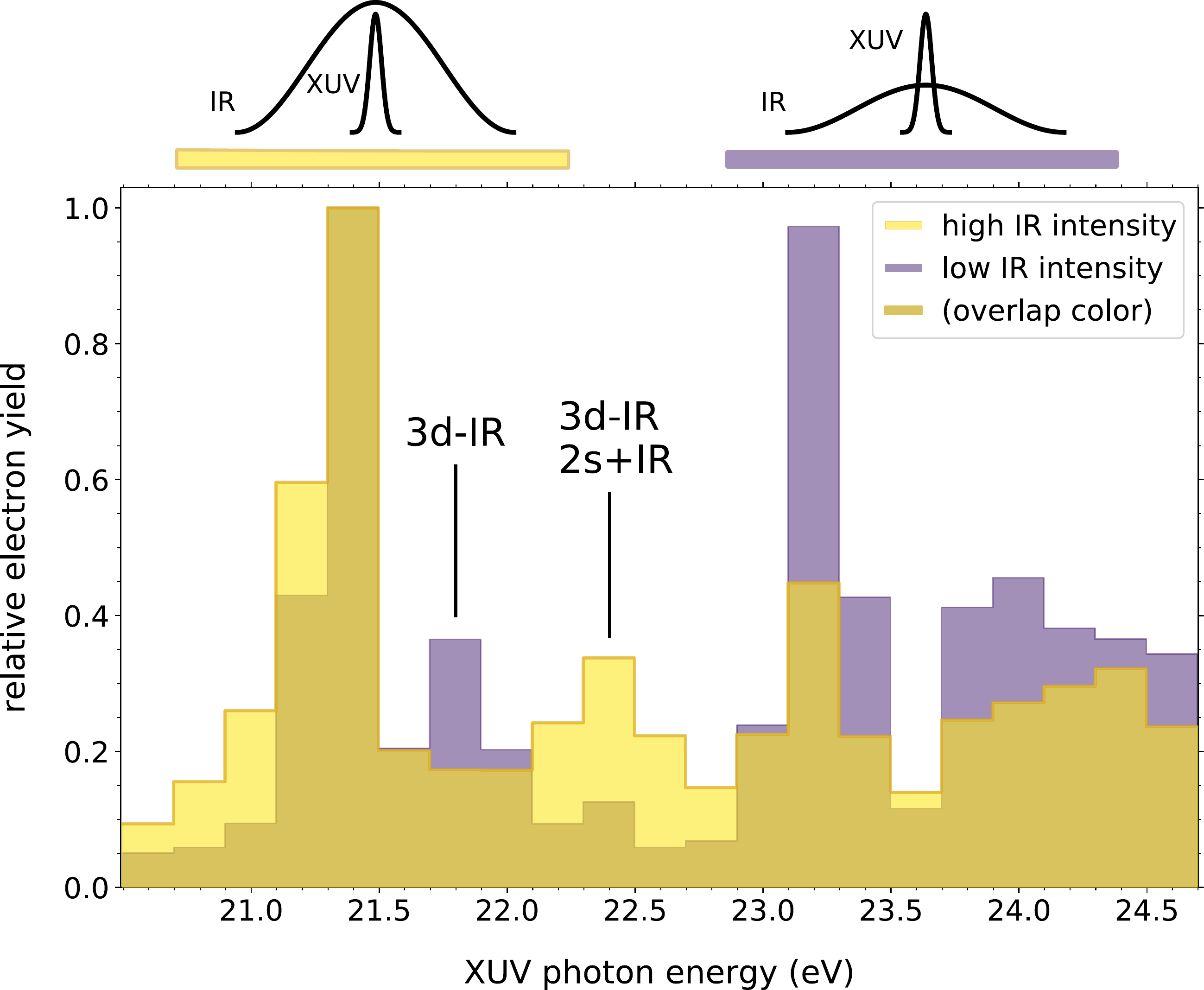}
	\caption{Photo\-electron yield measured for temporally overlapping XUV and IR pulses. 
		The distribution for high IR intensities ($I_{\rm high}$) is shown in yellow (light gray), 
		the distribution for low IR intensities ($I_{\rm low}$) in purple (dark). }
	\label{RelWeightInt}
\end{figure}

This mechanism is depicted in the lower part of \Cref{Levels}. Direct two-photon absorption 
couples the ground state to $1sns$ and $1snd$ states. In contrast to $P$-state ionization, 
peaks of LIS emerge at XUV energies matching the energy of the excited state plus/minus one IR photon.

In \Cref{RelWeightInt} the measured yield distributions for overlapping pulses 
are shown for the two cases of low and high IR intensity. 
The purple distribution in the background is recorded with $I_{\rm low}$, while the yellow distribution 
is again for the significantly larger intensity $I_{\rm high}$. The latter exhibits a gradual decrease in 
the yield from the peak corresponding to the $2p$ state at \SI{21.2}{\electronvolt} over the $3p$ state 
at \SI{23.2}{\electronvolt} up to the higher $np$ states (not visible due to the resolution). This overall 
decrease with rising XUV energy can be explained by the energy dependence of the cross section for the 
excitation step~\cite{DipTransMom}. The excitation probabilities are directly mapped to the ionization 
yield in the case of $I_{\rm high}$, where the IR intensity is large enough to ionize all excited states 
independent of the number of photons ($N$) needed. For low IR intensity (purple distribution in 
\Cref{RelWeightInt}), on the other hand, the scaling law $R \sim I^N$ of the transition rate $R$ with the 
laser intensity $I$ in multiphoton processes becomes relevant~\cite{AtomsBook}. 
Therefore, the ionization yield of the $2p$ state, which requires $N=3$ IR 
photons, is reduced compared to the yield from the $3p$ and higher $np$ states ($N=1$). 
  
Two approaches are employed to assign the LIS to specific field-free atomic excited states. First, we 
analyze the calculated distributions for excitation in combination with ionization and compare them 
with the experimental ionization yield. Second, the inspection of the measured photo\-electron angular 
distribution allows us to deduce the intermediate bound state that the electron was emitted from. 

\begin{figure}[b]
	\includegraphics[width=0.48\textwidth]{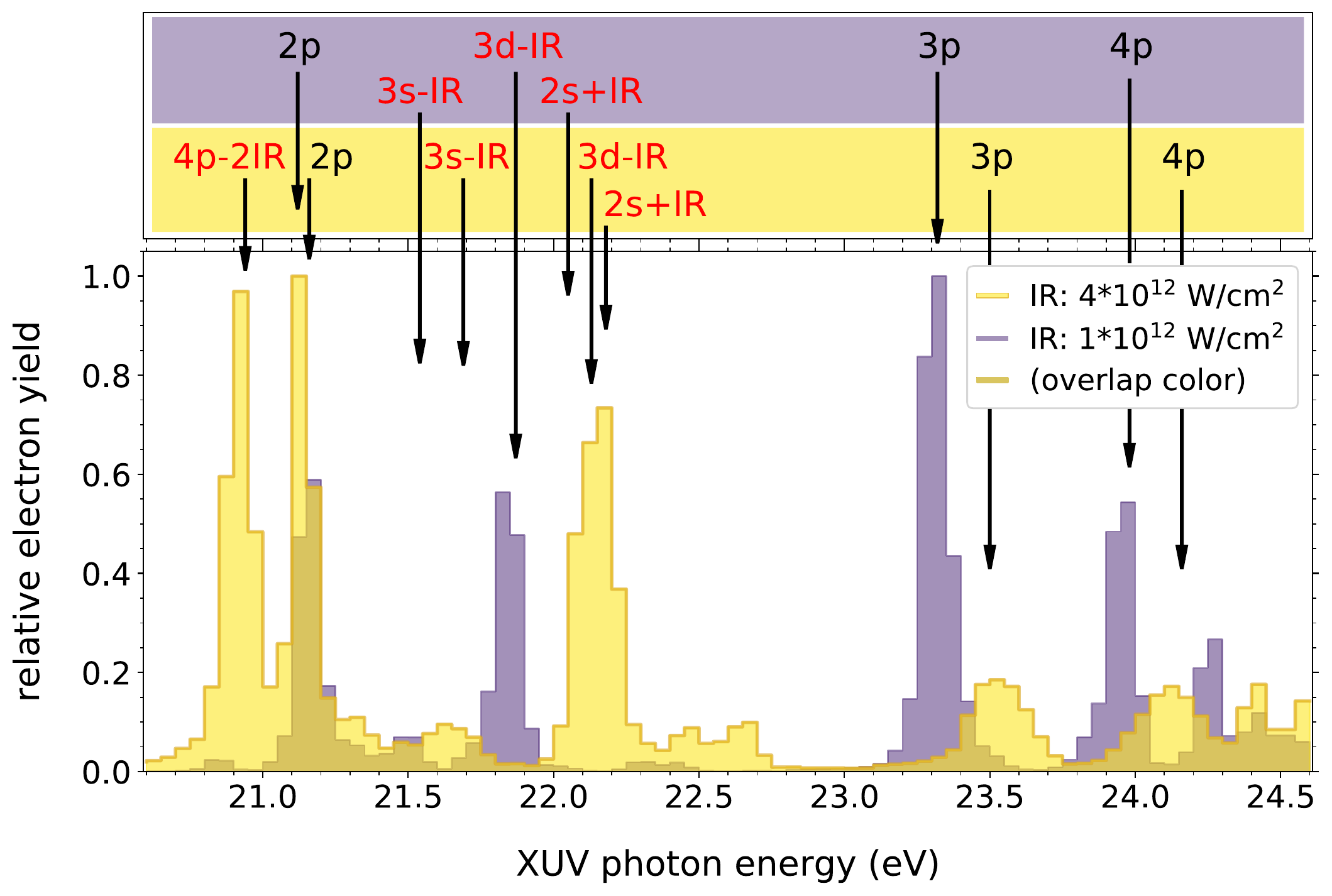}
	\caption{Calculated ionization probability for two different IR intensities. Arrows in the top 
		panel indicate the calculated positions of excited (black) and light-induced (red) states. 
		+\,(--) IR denotes the emission\,(absorption) of an IR photon.}
	\label{XUVyield_theo}
\end{figure}

\Cref{XUVyield_theo} shows theoretical predictions for XUV and IR pulses in temporal overlap. The 
calculated ionization probability is plotted against the XUV photon energy for two different IR 
intensities, color-coded in yellow and purple for high and low IR intensity, respectively (similar 
to the measurement shown in \Cref{RelWeightInt}). In addition to the ionization yield, our 
calculations predict the population distribution of excited atomic states at the end of the pulses 
as a function of the XUV energy and the IR intensity. The most prominently populated states in the 
theoretical excitation-probability distributions are marked by arrows in \Cref{XUVyield_theo}. 
They serve as an indicator of the role of the respective excited states and their population 
(either direct or light-induced) at a given XUV energy en route to ionization. This allows 
assignment of the peaks in the ionization probability distribution (bottom panel of 
\Cref{XUVyield_theo}) to the excited states from which the atom is ionized. 

Electronic energy levels experience an AC Stark shift due to the IR laser field~\cite{ACstark_1st,ACstark_Rev}. 
This shift is clearly seen for the $3p$ state in theory (see \Cref{XUVyield_theo}), but is
much smaller in the experiment (see \Cref{RelWeightInt}). However, experimental Stark/ponderomotive shifts can 
be seen in the photo\-electron kinetic energy spectrum (not shown here). Overall, we see good agreement 
between  experiment and theory when comparing the purple distributions in 
\Cref{RelWeightInt,XUVyield_theo}. Similar to the experiment, the calculated ionization yield shows a 
reduced contribution of the $2p$ state relative to the $3p$ state when the IR intensity is lowered. Moreover, 
the strengths and the positions of the LIS peaks, which only appear in temporal overlap, are well 
reproduced by the calculations. Small shifts in energy are attributed to the already mentioned 
inherent inaccuracies of electronic binding energies in the SAE model. Most importantly, for low IR intensity 
the dominating LIS peak is found around \SI{21.8}{\electronvolt} as in the experimental counter\-part. As the 
enhancements by other excited states appear at distinctively different positions, and are much less pronounced, 
this dominating LIS peak at \SI{21.8}{\electronvolt} (purple distribution in \Cref{RelWeightInt}) can be 
assigned to the $3d$ state.

The situation changes for high IR intensity where the dominant LIS peak is shifted to a larger XUV energy of
about \SI{22.4}{\electronvolt} in experiment (yellow in \Cref{RelWeightInt}). 
Comparison with theory indicates that in this case ionization proceeds through
the Stark-shifted $3d$ and $2s$ excited states. Both contribute to the dominating LIS peak at about 
\SI{22.2}{\electronvolt} according to our state assignment in \Cref{XUVyield_theo}. Relative to
the $3d$ state, the $2s$ contribution becomes more relevant at high intensity because, in order to populate 
the $2s$ state, the atom absorbs one XUV photon while emitting one IR photon. Ionization takes place by 
absorbing another three IR photons. In contrast to this effective four IR-photon transition, ionization 
via the $3d$ state involves only two IR photons. We note that the large shift of the $3d$ LIS with 
IR intensity seen in the experiment is also consistent with the calculation by Chen 
{\it et al.}~\cite{LISLeone2012}. 

\begin{figure}[]
	\includegraphics[width=0.5\textwidth]{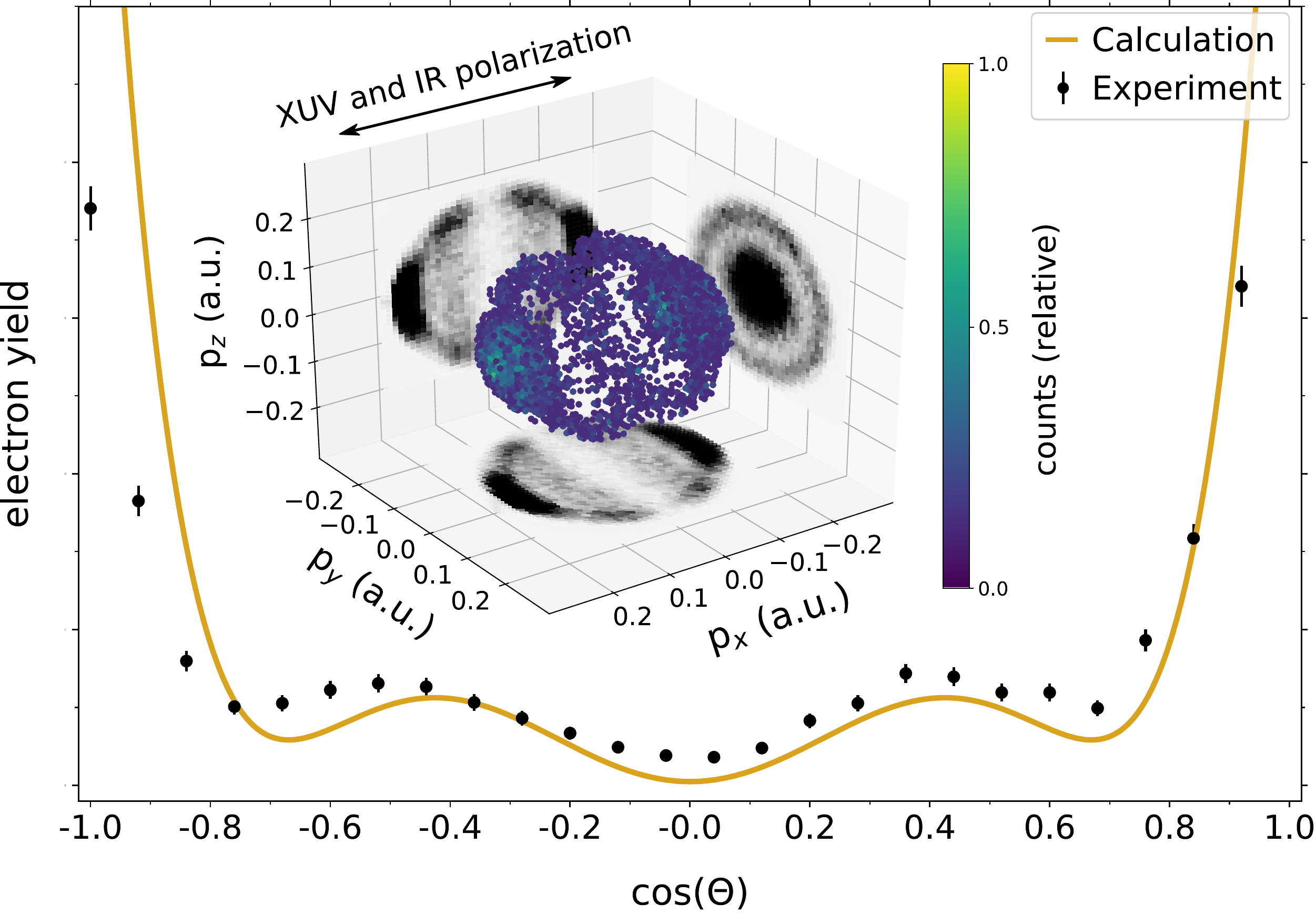}
	\caption{PAD for the dominant LIS at high IR intensity. 
		(Exp: $E_{\rm XUV}$ from \SIrange{22.2}{22.5}{\electronvolt} and $E_{\rm e}$ from 
		\SIrange{0.2}{0.5}{\electronvolt}, Calc: $E_{\rm XUV}$ from \SIrange{22}{22.3}{\electronvolt} and 
		$E_{\rm e}$ from \SIrange{0.2}{0.5}{\electronvolt}). The experimental counts are normalized to theory. 
		The inset shows the  corresponding measured 3D photo\-electron momentum distribution.}
	\label{AngularDistr}
\end{figure}

The electronic structure of the dominant LIS involving both the $2s$ and $3d$ excited states at 
\SI{22.4}{\electronvolt} for high IR intensity in \Cref{RelWeightInt} can also be deduced from the 
photo\-electron angular distribution (PAD). This is shown in \Cref{AngularDistr} where the yield of 
electrons is plotted as a function of their emission angle \(\Theta\) with respect to the laser 
polarization axis.

The distribution exhibits a typical \enquote{$F$-like} shape, indicating the angular quantum number of 
the continuum state to be \(L=3\). Starting from a \(^1\!S\) state (\(L=0\)), an angular quantum number of 
\(L=3\) can only be reached by absorbing at least three photons. For an atom in a \(^1\!D\) state (\(L=2\)), 
absorbing one IR photon is sufficient to obtain an $F$-like PAD. One can deduce the populated bound states by 
taking the photo\-electron kinetic energy (\SI{0.4}{\electronvolt}) minus the IR photon energies 
(each \SI{1.55}{\electronvolt}) while also accounting for the Stark-/ponderomotively shifted continuum level. 
One finds that the matching states are the $3d$ and $2s$ excited states. The yellow solid line in 
\Cref{AngularDistr} shows the calculated PAD for the considered LIS. We find very good agreement between 
experiment and theory, thereby supporting our interpretation of the mechanism.

The three-dimensional PAD of the dominant LIS is also contained in the measured three-dimensional 
photo\-electron momentum distribution, shown in the inset of \Cref{AngularDistr}. Dots in the plot 
represent a bin in momentum space, with the yield within each bin being color-coded. The energy range was 
chosen from \SI{0.3}{\electronvolt} to \SI{0.5}{\electronvolt}, resulting in a spherical shell of dots in 
the plot. The projections on the walls show the integrated yield along one specific direction. The two small 
maxima in \Cref{AngularDistr} are found as two rings around the polarization axis, while the large maxima 
are found in the three-dimensional distribution for maximal absolute \(p_x\) momentum, i.e. along the 
polarization axis of IR and XUV.


To summarize: We have measured and analyzed photo\-electrons stemming from laser-dressed atomic helium. 
The helium atoms were photo-excited over a large energy range by XUV FEL radiation in the 
presence of a moderately strong IR laser (\(\sim\!\SI{e12}{\watt\per\centi\meter\squared}\)). 
The scheme allowed us to reveal light-induced structures alongside the \(^1\!P\) Rydberg series. 
The observed excitation energies of the LIS for varying IR intensities 
were supported by TDSE calculations. The photo\-electron angular distribution and the kinetic energy 
were used to assign the LIS. The dominant LIS in the photo\-electron yield of 
laser-dressed helium was identified as stemming from the $1s3d\, ^1\!D$ excited state for an IR intensity 
of about \SI{1e12}{\watt\per\centi\meter\squared}, while for a higher IR intensity of 
about \SI{4e12}{\watt\per\centi\meter\squared} the $1s2s\,^1\!S$ state also contributes significantly.

Our investigation complements previous transient-absorption measurements on light-induced structures, 
but brings up additional aspects. While transient-absorption measurements reveal LIS in the 
absorption spectrum without the need for ionization, our measurement is sensitive to the excited 
states from which electrons are emitted and allows to measure and assign angular distributions of 
the corresponding continuum final states.

The presented analysis reveals the preference to populate the $1s3d\,^1\!D$ state rather than the 
$1s3s\,^1\!S$ state. This is cleary shown in the high IR intensity case, where all excited states 
get ionized, so the yield is independent of the number of ionizing photons. We find the propensity 
of helium in its ground state to increase angular momentum by two-photon absorption, thus driving 
a bound-bound transition. This can be understood in the context of Fano's propensity 
rule~\cite{FanoPropRule}, originally stated between bound and continuum states, and the propensity 
analysis of continuum-continuum transitions by Busto {\it et al.}~\cite{BustoPropRule}, both 
stating the propensity to increase the angular momentum in photo-absorption.


\medskip
The work was supported, in part, by the United States National Science Foundation under
grant Nos.~PHY-1803844 (KB) and PHY-2012078 (ND), and by the
XSEDE supercomputer allocation No.~PHY-090031 (AB,KB,ND). The
calculations were carried out on Comet at the San Diego Supercomputer Center and 
Frontera at the Texas Advanced Computing Center. AB is grateful for a Michael Smith Scholarship.

SA has received funding from the European Union’s Horizon 2020 research and innovation programme 
under the Marie Skłodowska-Curie grant agreement No. 701647.

\bibliography{references} 

\end{document}